%% file: main.tex
\documentclass[acmsmall,screen]{acmart}

\usepackage{amsmath,amsfonts,amsthm}
\usepackage{algorithmic}
\usepackage{graphicx}
\usepackage{textcomp}
\usepackage{xcolor}
\usepackage{booktabs}
\usepackage{tabularx}
\usepackage{array}
\usepackage{ragged2e}
\usepackage[most]{tcolorbox}
\usepackage{url}

\newcommand{\bad}[1]{\textcolor{red}{#1}}
\newcommand{\badcomment}[1]{\textcolor{red}{\normalfont\itshape #1}}

\AtBeginDocument{%
  }

\setcopyright{acmlicensed}
\copyrightyear{2018}
\acmYear{2018}
\acmDOI{XXXXXXX.XXXXXXX}
\acmConference[Conference acronym 'XX]{Make sure to enter the correct
  conference title from your rights confirmation email}{June 03--05,
  2018}{Woodstock, NY}
\acmISBN{978-1-4503-XXXX-X/2018/06}

\begin{document}

\title{Formal Model Construction Guided by Model-Based Proof Sketches}


\author{Hongshu Wang}
\email{hongshu.wang@u.nus.edu}
\affiliation{%
  \institution{National University of Singapore}
  \city{Singapore}
  \country{Singapore}
}

\author{Xinyue Zuo}
\email{zuoxy@nus.edu.sg}
\affiliation{%
  \institution{National University of Singapore}
  \city{Singapore}
  \country{Singapore}
}

\author{Yufan Cai}
\email{caiyf@nus.edu.sg}
\affiliation{%
  \institution{National University of Singapore}
  \city{Singapore}
  \country{Singapore}
}

\author{Neeraj Kumar Singh}
\email{neeraj.singh@toulouse-inp.fr}
\affiliation{%
  \institution{IRIT - National Polytechnic Institute of Toulouse}
  \city{Toulouse}
  \country{France}
}

\author{Yamine Ait Ameur}
\email{yamine.aitameur@toulouse-inp.fr}
\affiliation{%
  \institution{IRIT - National Polytechnic Institute of Toulouse}
  \city{Toulouse}
  \country{France}
}

\author{Jin Song Dong}
\email{dcsdjs@nus.edu.sg}
\affiliation{%
  \institution{National University of Singapore}
  \city{Singapore}
  \country{Singapore}
}



\begin{abstract}




  Formal modeling provides strong guarantees about system correctness, but developing and repairing formal models remains labor-intensive and requires substantial expertise in logic and formal reasoning.
Recent LLM-based autoformalization agents seek to reduce this burden by generating candidate formal models and revising them using feedback from formal tools.
However, the existing approaches follow a generate-and-repair paradigm, in which repairs are driven by verification failures of the generated model and therefore depend heavily on both the granularity of the feedback and the LLM's repair capability.
As a consequence, a repair targeting one level of verification may invalidate properties at another level, which requires reasoning over the complete set of event guards.
To address these limitations, we propose Proof-Sketch-Guided Formal Model Synthesis (\textit{ProGS}), an autoformalization method centered on model-based proof sketches.
A model-based proof sketch represents the proof structure of the target formal system as a tree.
Internal nodes capture case splits and inductive reasoning steps, while leaf nodes correspond to concrete state-transition events that realize individual subgoals.
\textit{ProGS} uses LLMs to generate and repair these sketches, with verification failures mapped back to specific nodes and subtrees to provide structured guidance for iterative repair.
Our evaluation on a benchmark of 27 formal systems shows that \textit{ProGS} improves over state-of-the-art agentic formal modeling approaches in syntactic validity, deductive verifiability, and behavioral correctness, demonstrating the benefit of organizing formal model construction around hierarchical proof sketches.
\end{abstract}

\ccsdesc[500]{Software and its engineering~Software verification and validation}
\ccsdesc[100]{Software and its engineering~Formal language definitions}

\keywords{Autoformalization, Intermediate Representation, Formal Verification, LLM Agent}

\received{20 February 2007}
\received[revised]{12 March 2009}
\received[accepted]{5 June 2009}

\maketitle

\input{1.intro}
\input{2.motivation}

\input{3.method}

\input{4.eval}

\input{5.discussion}

\input{6.related_work}

\input{7.conclusion}

\bibliographystyle{ACM-Reference-Format}
\bibliography{refs}

\appendix

\end{document}

%% file: 1.intro.tex
\section{Introduction}

Formal modeling provides a rigorous means for specifying system behavior and establishing correctness properties through model checking and theorem proving.
It has been widely applied to systems where failures can be costly, including concurrent protocols, safety-critical software, and distributed systems~\cite{zhang2025position}.
Despite these benefits, constructing a correct formal model remains labor-intensive.
Informal requirements must be translated into mathematical models that define system features, such as state variables and state-transition events, as well as correctness properties, such as invariants and reachability constraints.
When verification fails, the model must be repeatedly diagnosed and repaired.
These activities require substantial expertise in logic, specification languages, and formal reasoning, limiting the accessibility and scalability of formal modeling.
Recent advances in large language models (LLMs) have stimulated growing interest in automating formal model construction.
Existing LLM-based formal modeling approaches reduce manual effort by generating candidate formal models and using feedback from model checkers or theorem provers to diagnose and repair them~\cite{zuo2025pat,chengsysmobench, specula2025,alhanahnah2025empirical,hasan2026automated,chen2026modelwisdom,wang2026event}.
However, the effectiveness of this generate-and-repair paradigm depends on both the granularity of verification feedback and the LLM's ability to infer suitable repairs.
Such feedback is typically reported either at the level of the complete model or at the level of individual components, such as state-transition events.
Model-level feedback, such as a deadlock trace reported by a model checker, exposes violations of global properties but often provides limited guidance on which model components jointly contribute to the fault.
By contrast, component-level feedback, such as invariant-preservation proof obligations for individual state-transition events, can localize failures more precisely.
Nevertheless, existing approaches typically represent a formal model as a flat collection of generated components.
As a result, component-level feedback is often used to repair individual components in isolation, which can lead to inconsistent fixes, unexpected deadlocks, and models whose overall reasoning structure is difficult to interpret.
This mismatch between verification and repair granularity can cause a repair targeting one level of verification to invalidate properties checked at another level.
Therefore, the key challenge in system-level formal modeling is not only to generate or repair individual model components, but also to organize their dependencies so that local repairs remain consistent with global verification requirements.



To address this challenge, we propose \textbf{Pro}of-Sketch-\textbf{G}uided Formal Model \textbf{S}ynthesis (\textit{ProGS}), an autoformalization method that uses \textbf{model-based proof sketches} to organize dependencies among model components, localize faults precisely, and guide state-based formal model construction and repair.
A model-based proof sketch serves as an intermediate representation of the model under construction, connecting the hierarchical proof structure of the target system with the state-transition events that implement its behavior.
It represents the system as a tree, in which the internal nodes capture case splits and inductive reasoning steps, while leaf nodes correspond to transition events associated with individual proof subgoals.
This structure enables a divide-and-conquer construction process in which system-level reasoning is decomposed into node-level verification checks with simpler proof obligations, thereby reducing verification effort.
By localizing verification failures to specific nodes and grouping related failures through subtrees, \textit{ProGS} provides structured guidance for consistently repairing related model components, thereby reducing the modeling effort required to obtain a correct formal model.
In addition, this representation improves explainability by making the dependencies among model components explicit and allowing the effects of repairs to be traced across the model.
Given a natural-language system description, \textit{ProGS} first constructs an environment model, then generates and validates a model-based proof sketch, and finally synthesizes a formal model from the validated sketch.
During this process, \textit{ProGS} iteratively repairs the sketch using feedback from structural checks over sketch edges and proof obligations associated with leaf nodes.


We evaluate \textit{ProGS} on a benchmark of 27 formal systems and compare it with state-of-the-art agentic formal modeling approaches.
The results show that \textit{ProGS} achieves high proof-level verifiability while maintaining competitive runtime performance.
In contrast, the baselines tend to improve one aspect of performance at the cost of another.
Our efficiency analysis further shows that \textit{ProGS} reduces modeling effort compared with existing approaches that incorporate proof information, resolving proof failures with fewer LLM requests.
An ablation study confirms that both structural checks and subtree-based repair contribute to the overall performance.
These findings suggest that organizing formal model construction around model-based proof sketches provides more effective verification feedback and a more appropriate repair granularity than flat generate-and-repair workflows.

In summary, we make the following contributions:
\begin{itemize}
    \item We introduce model-based proof sketches, a hierarchical intermediate representation that connects proof decomposition to concrete state-transition events in the system model.
    
    \item We propose \textit{ProGS}, a novel LLM-based autoformalization method centered on model-based proof sketches that simplifies verification, enables more precise fault diagnosis, and provides structured guidance for formal model construction. We integrate \textit{ProGS} into the Rodin IDE~\cite{rodin} to demonstrate its effectiveness.
    
    \item We evaluate \textit{ProGS} on 27 formal systems and show that it outperforms existing agentic formal modeling approaches in syntactic validity, deductive verifiability, and runtime performance.

\end{itemize}



%% file: 2.motivation.tex
\section{Background and Motivation}

\subsection{Preliminary} \label{sec:preliminary}

\subsubsection{Formal Model}

A state-based formal model $M=(\Gamma, E)$ consists of an environment model $\Gamma$ and a set of state-transition events $E$.
The environment $\Gamma$ contains constants, sets, axioms, variables, invariants, and variants.
Each event $e=(grd, act) \in E$ consists of a guard $grd$ and a state-transition action $act$.
Given a before-state $s$ and an after-state $s'$, the guard $grd(s)$ is interpreted as a predicate specifying whether event $e$ is enabled in $s$; we write $grd$ when the state is clear from context.
The action $act(s, s')$ is interpreted as a transition predicate over the variables in $s$ and $s'$.
We use $I$ and $I'$ to denote the global invariants evaluated in the before-state and after-state, respectively.

\subsubsection{Verification Obligations}

For a model with $n$ events, deadlock-freeness among the reachable states requires that at least one event is enabled in every reachable state.
Formally,
\begin{equation}\Gamma \vdash \forall s. Reachable(s) \Rightarrow \bigvee_{i=1}^{n} grd_i(s)\label{eq:deadlockfree}
\end{equation}

Consider a global invariant $I$ that needs to be preserved by all $n$ events in the model.
The invariant preservation proof obligation can be expressed by Equation~\ref{eq:inv_preservation}.
\begin{equation}\Gamma \vdash \bigwedge_{i=1}^{n} (grd_i \land act_i \land I \Rightarrow I')\label{eq:inv_preservation}
\end{equation}







\subsubsection{Modeling Efforts}

Constructing a correct formal model involves both modeling effort and verification effort.
Modeling effort refers to the amount of human and LLM intervention required to obtain a correct model and repair it based on formal verification results.
In practice, $I$ in Equation~\ref{eq:inv_preservation} is usually decomposed into $m$  conjunctive invariants, i.e., $I = \bigwedge_{j=1}^{m} inv_j$, with a set of less complex proof obligations: $\{\Gamma \land grd_i \land act_i \land inv_j \vdash inv_j' \mid 1 \leq i \leq n, 1 \leq j \leq m\}$.
This decomposition distributes invariant-preservation checking across individual event-invariant pairs.
As a result, repair can be localized to a specific event and invariant.
In contrast, Equation~\ref{eq:deadlockfree} cannot be decomposed into proof obligations of this form due to its disjunctive nature.
As a result, when such a verification obligation fails, the verification result provides limited localization information, making it difficult to identify which event or guard should be repaired.
Therefore, the modeling effort associated with deadlock-freeness can be substantial.


\subsection{Problem Statement}

We define state-based formal modeling as the task of constructing a model $M=(\Gamma, E)$.
The constructed model is expected to satisfy key correctness properties, including deadlock-freeness and invariant preservation.
Existing autoformalization approaches generate $M$ and repair it using verification feedback reported either at the model level or at the event level.
Model-level verification, such as checking deadlock-freeness through the disjunction of all event guards, detects violations of global properties but provides limited information for localizing the fault.
Event-level verification, such as invariant-preservation proof obligations, provides more localized feedback, but repairing failed events independently may ignore hidden dependencies among related events and introduce inconsistencies into the model.
Therefore, effective autoformalization should repair models at a more appropriate granularity.
We define a repair group $R\subseteq E$ as a set of events whose verification failures are likely to share a common structural cause.
Instead of repairing each failed event independently ($|R| = 1$) or repairing the entire model monolithically ($R=E$), an effective repair strategy should identify repair groups that are both sufficiently small to avoid unnecessary changes and sufficiently complete to cover the events affected by the same underlying fault.

\subsection{Motivating Example} \label{sec:motivation}

\input{tables/sorting_invariants}
\input{tables/motivation_baselines}

\begin{figure*}[t]
    \centering
    \includegraphics[width=1\textwidth]{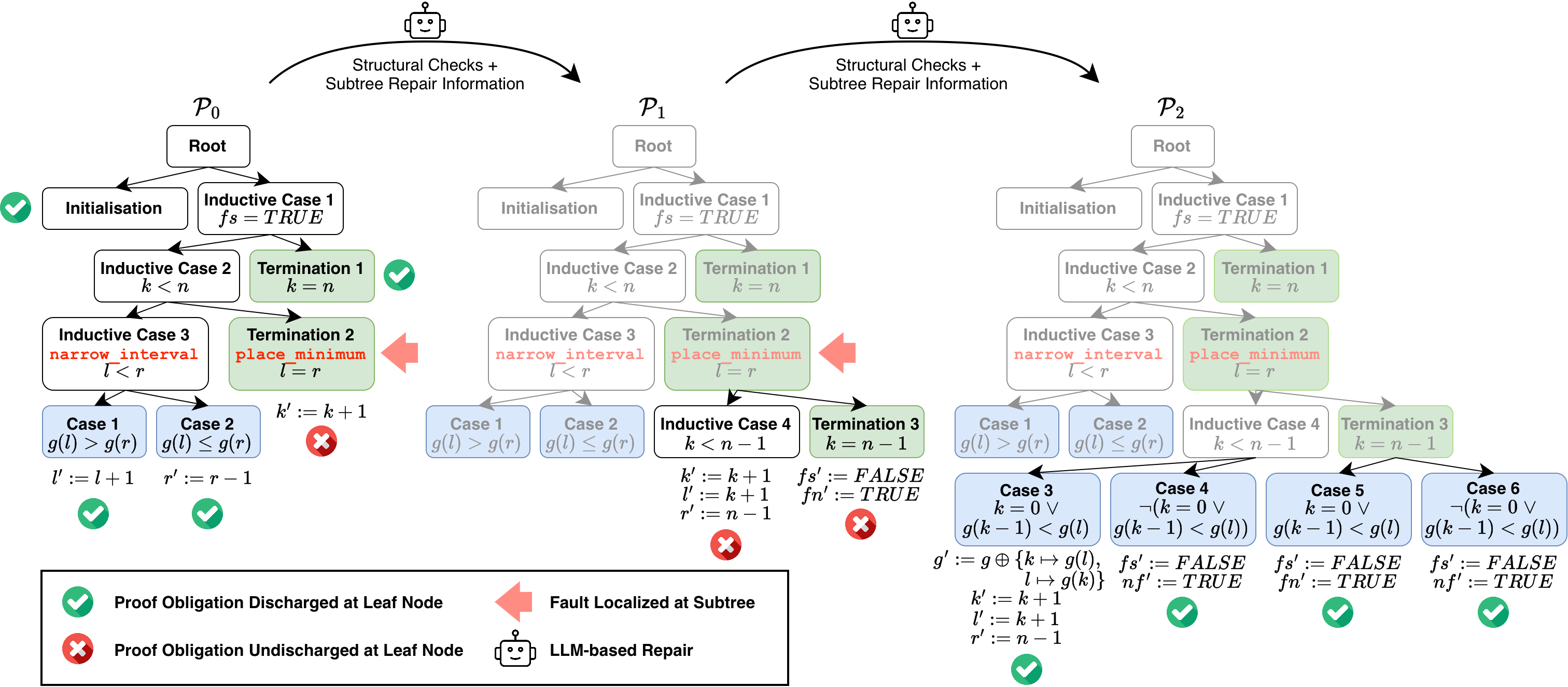}
    \caption{Model-based proof sketches generated by \textit{ProGS}. Green nodes denote termination cases, while blue nodes denote general cases introduced by case splitting. Each leaf node is annotated with its corresponding action. Each node is labeled with guards, where the common guards shared with its parent are omitted for simplicity.}
    \label{fig:motivation}
\end{figure*}


Existing approaches to system-level formal modeling often fail to construct coherent formal models and provide limited support for generating high-quality repairs from verification feedback.
Consider a selection sort system that sorts a finite sequence $g$ of size $n$ into ascending order.
During sorting, a boundary variable $k$ separates the sorted prefix from the unsorted suffix of the sequence, while scanning indices $l$ and $r$ locate the minimum element in the suffix, which is then swapped into the boundary position.
The core safety requirements of this system are summarized in Table~\ref{tab:nl_formal_req}.

State-of-the-art methods, including Basic Modeling Agent~\cite{chengsysmobench}, PAT-Agent~\cite{zuo2025pat}, and Event-B Agent~\cite{wang2026event}, struggle to produce a high-quality formal model for this sorting algorithm.
As illustrated in Figure~\ref{fig:motivation_baselines}(a), the state transition event \texttt{narrow\_interval} assumes the existence of new bounds $l1$ and $r1$ that satisfy the minimum interval containment property (SAF-3), without specifying how these bounds are computed.
The model then updates the current bounds $l$ and $r$ to be $l1$ and $r1$, respectively.
This results in a less interpretable model that relies on the $min$ function without explicitly verifying the correctness of the minimum searching procedure.

PAT-Agent follows a similar generate-and-repair procedure based on model checking.
Although the generated formal model is correct under bounded model checking, some events cannot be fully verified because the method does not perform event-level theorem proving.
In particular, the \texttt{place\_minimum} event in Figure~\ref{fig:motivation_baselines}(b) swaps the identified minimum value in the unsorted suffix of the sequence, $g(l)$, into the boundary position $k$.
However, the proof obligation to preserve prefix-sortedness (SAF-5) before and after the event cannot be discharged.
The failure is caused by a missing guard, $g(k-1) < g(l)$, which is needed for automated provers to establish that the updated prefix $g[0..(k'-1)]$ remains sorted after the swap.

Discovering such additional model components is non-trivial.
Event-B Agent incorporates proof results into the repair process, enabling it to add missing components to discharge failed proof obligations.
However, its generated model illustrates that repairing a model solely to satisfy a proof obligation can introduce semantic errors.
As shown in Figure~\ref{fig:motivation_baselines}(c), Event-B Agent adds the guards $l=k+1$ and $g(k+1)<g(k+2)$ to \texttt{place\_minimum}.
These guards over-constrain the event by requiring the minimum element of the unsorted suffix to already appear at its first position.
This effectively assumes that $g$ is already sorted before the sorting procedure begins, thereby introducing unexpected deadlocks.

\textit{ProGS} addresses the limitations of the baselines through verification and repair of model-based proof sketches.
As shown in Figure~\ref{fig:motivation}, \textit{ProGS} constructs a model-based proof sketch as an intermediate representation of the formal model, organizing state-transition events into a tree-like structure.
Given the constructed environment model, an LLM first generates an initial sketch $\mathcal{P}_0$.
\textit{ProGS} then iteratively refines the sketch using formal verification feedback, eventually producing the final sketch $\mathcal{P}_2$.
The baseline events \texttt{narrow\_interval} and \texttt{place\_minimum} correspond to labeled nodes in the model-based proof sketches.
In $\mathcal{P}_0$, \texttt{narrow\_interval} is decomposed into two child nodes (Cases 1 and 2), representing the two cases in the minimum-search procedure.
Unlike Basic Modeling Agent, which assumes the availability of a $min$ function, \textit{ProGS} generates verifiable events that explicitly model how the minimum is searched.

For \texttt{place\_minimum}, \textit{ProGS} initially generates a single node with an incomplete action set, $\{k' := k+1\}$, similar to PAT-Agent and Event-B Agent.
As a result, the proof obligation for preserving prefix sortedness cannot be discharged.
To repair this failure, \textit{ProGS} refines the single \texttt{place\_minimum} node in $\mathcal{P}_0$ into four nodes in $\mathcal{P}_2$.
These nodes cover the successful swap of the identified minimum (Case 3), two abnormal cases (Cases 4 and 6), and the successful termination of the algorithm (Case 5).
Unlike Event-B Agent, \textit{ProGS} identifies and adds the critical guard $grd_{critical}: k=0 \lor g(k-1) < g(l)$, while preserving behavioral coverage by also including cases where $\neg grd_{critical}$ holds.

%% file: tables/sorting_invariants.tex
\begin{table}[t]
\centering
\caption{A subset of natural-language safety requirements and their formalizations of the selection sort algorithm.}
\label{tab:nl_formal_req}
\scriptsize
\setlength{\tabcolsep}{2.5pt}
\renewcommand{\arraystretch}{1.12}

\begin{tabular}{@{}
p{0.06\columnwidth}
p{0.36\columnwidth}
p{0.52\columnwidth}
@{}}
\toprule
\textbf{ID} & \textbf{Requirement} & \textbf{Formalization} \\
\midrule

SAF-3 &
The candidate minimum is never lost during scanning: the minimum of the unsorted suffix of $g$ is always within the interval bounded by $l$ and $r$. &
\(\begin{array}[t]{@{}l@{}}
(fs = TRUE \land k \leq n - 1) \Rightarrow (min(g[k..(n - 1)]) \in g[l..r])
\end{array}\)
\\
\addlinespace[0.4em]

SAF-4 &
When $l$ equals $r$, the minimum of the unsorted suffix of $g$ should be located at this position. &
\(\begin{array}[t]{@{}l@{}}
(fs = TRUE \land k \leq n-1 \land l=r) \Rightarrow
(min(g[k..(n-1)]) = g(l))
\end{array}\)
\\
\addlinespace[0.4em]

SAF-5 &
While $fs = true$, the prefix of $g$ bounded by $k$ is always sorted in strictly increasing order. &
\(\begin{array}[t]{@{}l@{}}
(fs = TRUE) \Rightarrow \\
(\forall i,j \cdot (i \in 0..(k - 1) \land{} j \in 0..(k - 1) \land{} i < j \Rightarrow g(i) < g(j)))
\end{array}\)
\\
\addlinespace[0.4em]

SAF-6 &
When the algorithm terminates, the entire array $g$ is sorted in strictly increasing order. &
\(\begin{array}[t]{@{}l@{}}
(fn = TRUE) \Rightarrow \\
(\forall i,j \cdot (i \in 0..(n - 1) \land{} j \in 0..(n - 1) \land i < j \Rightarrow g(i) < g(j)))
\end{array}\)
\\

\bottomrule
\end{tabular}
\end{table}

%% file: tables/motivation_baselines.tex
\begin{figure}[t]
\centering
\scriptsize
\ttfamily

\makebox[\columnwidth][c]{%
\begin{minipage}[t]{0.31\columnwidth}
\hrule
\vspace{2pt}
\begin{tabbing}
xxxx\=xxxx\=xxxx\=xxxx\=\kill
\textbf{(a) Basic Modeling Agent} \\
transition narrow\_interval \\
parameter $l1, r1$ \\
guard \\
\> $fs = TRUE$ \\
\> $l \leq l1 \land l1 \leq r1 \land r1 \leq r$ \\
\> $r1 - l1 < r - l$ \\
\> \bad{$min(g[k..(n-1)]) \in g[l1..r1]$} \\
\> \badcomment{// unverified computation of bounds} \\
action \\
\> $l' = l1$ \\
\> $r' = r1$ \\
\end{tabbing}
\vspace{-6pt}
\hrule
\end{minipage}%
\hfill
\begin{minipage}[t]{0.31\columnwidth}
\hrule
\vspace{2pt}
\begin{tabbing}
xxxx\=xxxx\=xxxx\=xxxx\=\kill
\textbf{(b) PAT-Agent} \\
transition place\_minimum \\
guard \\
\> $fs = TRUE$ \\
\> $l = r$ \\
\> $k < n - 1$ \\
\> \textcolor{pink}{$g(k-1) < g(l)$} \\
\> \badcomment{// missing guard} \\
\> $\phantom{g(k + 1) < g(k + 2)}$ \\
action \\
\> $g' = g \oplus \{k \mapsto g(l), l \mapsto g(k)\}$ \\
\> $k' = k + 1$ \\
\> $l' = k + 1$ \\
\> $r' = n - 1$ \\
\end{tabbing}
\vspace{-6pt}
\hrule
\end{minipage}%
\hfill
\begin{minipage}[t]{0.31\columnwidth}
\hrule
\vspace{2pt}
\begin{tabbing}
xxxx\=xxxx\=xxxx\=xxxx\=\kill
\textbf{(c) Event-B Agent} \\
transition place\_minimum \\
guard \\
\> $fs = TRUE$ \\
\> $l = r$ \\
\> $k < n - 2$ \\
\> \bad{$l = k + 1$} \\
\> \bad{$g(k+1) < g(k+2)$} \\
\> \badcomment{// deadlock introduced} \\
action \\
\> $g' = g \oplus \{k + 1 \mapsto g(l), l \mapsto g(k + 1)\}$ \\
\> $k' = k + 1$ \\
\> $l' = k + 2$ \\
\> $r' = n - 1$ \\
\end{tabbing}
\vspace{-6pt}
\hrule
\end{minipage}%
}

\caption{State-transition events generated by (a) the Basic Modeling Agent, (b) PAT-Agent, and (c) Event-B Agent for the selection-sort example.}
\label{fig:motivation_baselines}
\end{figure}

%% file: 3.method.tex
\section{Methodology}

\subsection{Overview}
In this work, we propose \textit{ProGS}, \textbf{Pro}of-Sketch-\textbf{G}uided Formal Model \textbf{S}ynthesis, to address the challenges of automated state-based formal model construction.
Given a natural-language requirements document, \textit{ProGS} uses LLMs to construct an environment model and identify dependencies among invariants and variants.
It then prompts a dedicated LLM to generate an initial model-based proof sketch.
The sketch structure is iteratively refined using additional checks for deadlock-freeness and parent-child guard validity.
Leaf nodes in the proof sketch are translated into concrete state-transition events and verified through event-level proof obligations.
Finally, the verification failures are mapped back to the proof sketch, localized at the subtree level, and provided to the LLM so that repairs can be focused on the problematic subtrees.



\subsection{Environment Model Construction}

As introduced in Section~\ref{sec:preliminary}, the environment model $\Gamma$ provides the foundation for formal model verification.
According to the system requirements document, \textit{ProGS} first constructs a high-quality environment model that accurately formalizes essential elements, including invariants and variants.
It prompts an LLM to generate the environment model and then iteratively repairs compilation and typing errors until the model is well-formed.
For the sorting example, the environment model includes the constants and variables described in Section~\ref{sec:motivation}, together with invariants and variants, some of which are shown in Table~\ref{tab:nl_formal_req}.

Requirements are formalized as invariants of the form $inv: condition \Rightarrow property$, where the condition may be $TRUE$, indicating that the requirement is formalized as a global invariant.
An example invariant in the selection sort example is the prefix-sortedness requirement (SAF-5), which can be formalized as $(fs=TRUE) \Rightarrow (\forall i,j \cdot (i \in 0..(k-1) \land j \in 0..(k-1) \land i < j \Rightarrow g(i) < g(j)))$.
Here, the condition $fs=TRUE$ specifies the scenario in which the property is expected to hold, allowing loop invariants and postconditions to be expressed.


After obtaining a well-formed environment model, \textit{ProGS} identifies dependencies among invariants and variants using an LLM, and then validates the proposed dependencies through formal checks.
These dependencies guide the construction of model-based proof sketches.
\textit{ProGS} considers four types of dependencies.
\textbf{Equivalence} and \textbf{Implication} identify duplicated or subsumed invariants.
\textbf{Case Split} determines whether a logical property, such as an invariant, can be decomposed into multiple cases, thereby suggesting the branching structure of the model-based proof sketch.
\textbf{Induction} identifies a relation among a candidate induction hypothesis, an induction conclusion, and a decreasing variant.
The variant defines the progress toward a base or terminal case in which the induction conclusion should hold.
The induction hypothesis is then used to establish the conclusion in the inductive step.

The dependencies are validated using the following proof obligations, where $P$, $Q$, and $Case_i$ denote the $property$ part of invariants:
\begin{enumerate}
    \item \textbf{Equivalence}: $P \Leftrightarrow Q$.
    \item \textbf{Implication}: $P \Rightarrow Q \land \lnot(Q \Rightarrow P)$.
    \item \textbf{Case Split}: $P \Leftrightarrow \bigvee_{i=1}^{n} Case_i$.
    \item \textbf{Induction}: Given $inv_1: condition_1 \Rightarrow P$ and $inv_2: condition_2 \Rightarrow Q$, \textit{ProGS} checks whether $P$ can serve as an inductive hypothesis for establishing $Q$ at the terminal variant value $var=0$.
    This dependency is validated by checking $\lnot(P \Rightarrow Q) \land \lnot(var=0 \Rightarrow Q) \land (P \land var=0 \Rightarrow Q)$.
\end{enumerate}

In the sorting example, \textit{ProGS} identifies $n-k$ as a variant.
When this variant reaches 0, the sorted-prefix invariant implies the sortedness of the entire array:
\[
\begin{aligned}
& (\forall i,j \cdot (i \in 0..(k-1) \land j \in 0..(k-1) \land i < j \Rightarrow g(i) < g(j)) \land \; (n-k=0) \Rightarrow \\
& (\forall i,j \cdot (i \in 0..(n-1) \land j \in 0..(n-1) \land i < j \Rightarrow g(i) < g(j)).
\end{aligned}
\]
This dependency suggests an inductive argument in which the sortedness of the prefix serves as the induction hypothesis, while the variant $n-k$ captures progress toward the final sorted state.
Similarly, $r-l$ is identified as the variant that connects SAF-3 and SAF-4, representing progress towards terminating the inner minimum-searching loop in the selection sort procedure.

\subsection{Model-Based Proof Sketch} \label{sec:sketch}

A model-based proof sketch is defined as a tree $\mathcal{P} = (\mathcal{N}, \mathcal{E})$, where $\mathcal{N}$ is a finite set of nodes, and $\mathcal{E} \subseteq \mathcal{N} \times \mathcal{N}$ is a finite set of directed edges.
Each leaf node $n_l=(grd,act) \in \mathcal{N}$ consists of guards $grd$ and actions $act$, and corresponds to a concrete state-transition event.
In contrast, each internal node $n_i = (grd, \{\}) \in \mathcal{N}$ contains guards but no concrete actions, representing an intermediate proof or case-splitting condition.
An edge $(n_p, n_c) \in \mathcal{E}$ connects a parent node $n_p$ to a child node $n_c$.
\textit{ProGS} first uses an LLM to generate an initial model-based proof sketch $\mathcal{P}$ based on the set of requirements, environment model $\Gamma$, and invariant dependencies.
After this, $\mathcal{P}$ is deterministically cleaned and iteratively repaired based on the results of structural checks.

\subsubsection{Sketch Cleaning}

LLMs may generate ill-formed or redundant structures in a model-based proof sketch.
Therefore, after generating the initial sketch, \textit{ProGS} applies deterministic normalization rules before performing formal structural checks:
\begin{itemize}
    \item \textbf{Removing spurious leaves.}
    Leaf nodes without actions are removed unless they explicitly represent terminal, observational, or skip events.
    This prevents non-executable leaves from being translated into unintended state-transition events.

    \item \textbf{Merging unary branches.}
    If an internal node has only one child, the child is merged into the parent to simplify the model-based proof sketch.
    The guards of the two nodes are combined to preserve the corresponding path condition.

    \item \textbf{Factoring common guards.}
    If multiple sibling nodes share common guards, \textit{ProGS} introduces an abstract parent node that contains the shared guards and retains only branch-specific guards in the children.
    This makes the subtree structure more explicit, thereby supporting hierarchical fault localization and repair in later stages.
    
\end{itemize}


\subsubsection{Structural Checks of a Model-Based Proof Sketch}

In addition to deterministic normalization, \textit{ProGS} performs formal structural checks in $\Gamma$ to ensure that a model-based proof sketch is well organized before it is translated into a formal model.
Specifically, two structural properties must hold.
Deadlock-freeness ensures that each parent node is fully covered by its child branches, while parent-child guard validity ensures that each child branch remains consistent with the condition represented by its parent.

\textbf{- Deadlock-Freeness}
\begin{equation}
\Gamma \vdash \bigwedge_{n_i \in \mathcal{N}} \left(n_i.grd \Rightarrow \bigvee_{{n_c \mid (n_i,n_c) \in \mathcal{E}}} n_c.grd \right)
\label{eq:deadlockfree_proof_sketch}
\end{equation}

Equation~\ref{eq:deadlockfree_proof_sketch} defines the deadlock-freeness check over a model-based proof sketch.
For each parent internal node $n_i$, its guards $n_i.grd$ must imply the disjunction of the guards of its child nodes.
This ensures that every case represented by a parent node is covered by at least one child branch, thereby preventing case splitting from introducing unexpected deadlocks.
Since the obligation is expressed as a conjunction over sketch nodes, it can be decomposed into smaller proof obligations, allowing verification at the edge-level.

By enforcing this coverage condition at every case split, the discharged local obligations collectively establish the overall deadlock-freeness property in Equation~\ref{eq:deadlockfree}.



\textbf{- Parent-Child Guard Validity}
\begin{equation}
\Gamma \vdash \bigwedge_{(n_p,n_c) \in \mathcal{E}} (n_c.grd \Rightarrow n_p.grd)
\label{eq:parent_child_guard}
\end{equation}

Equation~\ref{eq:parent_child_guard} defines the parent-child guard validity check.
For every edge $(n_p,n_c)$ in $\mathcal{P}$, the guards of the child node 
$n_c$ must imply the guards of its parent node $n_p$.
This ensures that each child branch represents a valid refinement or specialization of the condition captured by its parent.

Together with the coverage condition in Equation~\ref{eq:deadlockfree_proof_sketch}, this validity check ensures that the child branches exactly cover the behavior represented by their parent.
Every internal parent node is covered by at least one child branch, and every child branch remains within the parent condition.
Therefore, if an invariant is preserved by all child branches, then the invariant is preserved by the subtree with the parent node as the root.
Since invariant-preservation proof obligations are discharged at the leaf nodes, this preservation argument can be propagated upward through the proof sketch, eventually establishing that the invariant is preserved by the complete system represented at the root.

Both Equation~\ref{eq:deadlockfree_proof_sketch} and Equation~\ref{eq:parent_child_guard} can be decomposed into simpler proof obligations.
In the selection sort example shown in Figure~\ref{fig:motivation}, after the initial model-based proof sketch is generated by the LLM and deterministically cleaned, $\mathcal{P}_0$ is obtained.
An example deadlock-freeness check on $\mathcal{P}_0$ is $\Gamma \vdash fs=TRUE \Rightarrow k<n \lor k=n$, which holds because $k \leq n$ is included in $\Gamma$.
The corresponding child nodes must have guards $fs=TRUE \land k<n$ and $fs = TRUE \land k=n$ to fulfill the parent-child guard validity check.

\subsubsection{Model-Based Proof Sketch Structural Cost}

Based on the two structural checks, \textit{ProGS} computes a Structural Cost Score for $\mathcal{P}$:
\[
s_{cost}=\frac{1}{3} \times (\frac{|structural\_check\_fails|}{|\mathcal{N}|} + \frac{1}{depth} + \frac{1}{|leaf\_nodes|}),
\]
This score estimates the structural quality of the generated proof sketch.
A lower score is preferred, as it indicates fewer structural check violations, a deeper sketch that can support more precise fault localization, and more leaf nodes that decompose the model into finer-grained cases to ease proof obligation discharge.
Using a best-of-$n$ strategy with $n=3$, \textit{ProGS} generates three candidate proof sketches and selects the one with the lowest structural cost score.

The selected sketch and the failed structural checks are then provided to the LLM for iterative repair.
This process continues until the sketch satisfies all structural checks or a predefined trial limit is reached.

\subsection{Formal Model Synthesis and Subtree-Based Repair} \label{sec:subtree_repair}

\subsubsection{Formal Model Construction}

After the initial model-based proof sketch is constructed and validated by the structural checks, \textit{ProGS} derives a formal model from it.
Each leaf node $n_l \in \mathcal{N}$ in $\mathcal{P}$ is translated into a state-transition event $(n_l.grd,n_l.act) \in E$.
The events are then combined with the environment model $\Gamma$ to form the complete formal model $M$.

For example, in Figure~\ref{fig:motivation}, nodes Case 1 and Case 2 are converted into two state-transition events for searching the minimum in the unsorted suffix of sequence $g$.
While node \texttt{place\_minimum} represents an event to swap the minimum into the desired position.
Specifically, Case 1 corresponds to an event with guard $fs=TRUE \land k < n \land l < r \land g(l) > g(r)$, and action $l' := l+1$.

\subsubsection{Failure Localization and Subtree-Based Repair}

After constructing the formal model, \textit{ProGS} generates proof obligations at the event level, including well-definedness of guards, invariant preservation, and variant decrease.
Automated provers and SMT solvers are then applied to discharge these proof obligations.

For each invariant or variant with failed proof obligations, \textit{ProGS} maps the failures back from formal model events to the corresponding leaf nodes in the proof sketch.
Given a set of failed event nodes, \textit{ProGS} identifies subtrees whose descendant leaves all fail-proof obligations associated with the same invariant or variant.
These subtrees indicate that the failure may originate from a common structural condition, such as shared guards introduced by an internal node.
Therefore, instead of repairing each failed event independently, \textit{ProGS} prompts the LLM to repair the proof sketch at the subtree level, enabling coordinated changes to related events.

Figure~\ref{fig:motivation} demonstrates examples of failure localization in $\mathcal{P}_0$ and $\mathcal{P}_1$.
In $\mathcal{P}_0$, node \texttt{place\_minimum} fails to discharge the proof obligation on prefix-sortedness.
This information is provided to the LLM to generate a refined sketch $\mathcal{P}_1$, with case splits under node \texttt{place\_minimum} to attempt to discharge the proof obligation.
Since both leaf nodes still fail to discharge the proofs, the subtree \texttt{place\_minimum} is located, with all of its descendant nodes failing the same proof obligation.
The LLM is then instructed to repair the subtree \texttt{place\_minimum} as a whole, which corresponds to the set of events $R = \{\text{Inductive Case 4}, \text{Termination 3}\}$ to be repaired, yielding a final sketch $\mathcal{P}_2$.




\textit{ProGS} repeats structural check and subtree-based repair until it obtains a model-based proof sketch that is structurally valid and whose derived formal model discharges all generated proof obligations.
The validated sketch can then be deterministically translated into a set of state-transition events that avoid unexpected deadlocks and satisfy the user-defined properties encoded as invariants.

%% file: 4.eval.tex
\section{Evaluation}

We evaluate \textit{ProGS} from three perspectives: the quality of the generated formal models, the efficiency and modeling efforts of the construction process, and the contribution of individual components.

Specifically, we investigate the following research questions:
\begin{itemize}
    \item \textbf{RQ1.} How effective is \textit{ProGS} in constructing system-level formal models?
    \item \textbf{RQ2.} How efficient is \textit{ProGS}, and to what extent does it reduce modeling effort?
    \item \textbf{RQ3.} How does each component of \textsc{ProGS} contribute to its overall performance?
\end{itemize}

\subsection{Experimental Setup and Implementation} \label{sec:experiment_setup}

We compare \textit{ProGS} against the existing formal modeling approaches to evaluate its effectiveness.
The baselines specialize in various formal specification languages for state-transition systems, including Event-B~\cite{abrial2010modeling}, TLA+~\cite{lamport2002specifying}, PAT~\cite{sun2009pat}.
To ensure fairness in evaluation, we adopt Event-B as the common specification language, since it supports all the formal verification tools used in the baselines, eases proof obligation generation, and is closely related to proof by induction.
In contrast, TLA+ and PAT are mainly designed for model checking without proof decomposition as described in Section~\ref{sec:preliminary}.

Specifically, to adapt the baselines and metrics to our experiments, the Rodin IDE~\cite{rodin} for Event-B is used to report compilation errors, the automated provers and SMT solvers supported by Rodin are used to discharge proofs, and ProB~\cite{ProB} is used as the model checker.
In addition, the baselines use syntax and rules specific to the formal languages they adopt to optimize performance.
We replace such information with the Event-B syntax documentation and rules to preserve the baselines' capabilities while ensuring a fair comparison across methods.

Throughout the experiments, GPT 5.4 mini~\cite{openai2026gpt54mini} is used as the large language model for all methods and metric computations.


\subsection{Benchmarks}

We evaluated \textit{ProGS} and the baselines on a benchmark of 27 examples from Event-B Agent~\cite{wang2026event}, covering sequential algorithms, concurrent systems, and distributed systems.

The datasets were manually processed to ensure a consistent input format.
The input for each example in the dataset is a natural-language requirements document containing equipment (EQP), functional (FUN), and safety (SAF) requirements.
The EQP requirements describe the name, type, and purpose of each constant, and any additional types to be used by the system.
The constant information is given in the requirement document to enable the computation of runtime coverage and trace conformance metrics, since both metrics take predefined constants and values as inputs.
For similar reasons, the names, types, and purposes of the variables are specified in the FUN requirements.
The FUN requirements also describe the expected system behavior in natural language, without detailing how the variables should be updated under different conditions.
Lastly, the SAF requirements describe the properties that the system should maintain, including loop invariants, postconditions, and global invariants.
Compared with the FUN requirements, they are specified in greater detail, allowing them to be translated into formal invariants with minimal variation across methods and thereby supporting a fairer comparison.
Some examples of natural language requirements can be found in Table~\ref{tab:nl_formal_req}.
Detailed information is provided to enable the computation of proof-related metrics, rather than to provide guidance on constructing state transition events.

For each system in the dataset, we prepare three test cases, each defined by a concrete assignment of constant values, and use ProB to collect execution traces from the ground-truth formal models.
These traces are then used to compute the runtime-performance metrics.

\subsection{Baselines}
\textit{ProGS} is compared against 3 baselines, Basic Modeling Agent, PAT-Agent, and Event-B Agent with the adaptation described in Section~\ref{sec:experiment_setup}.
Basic Modeling Agent~\cite{chengsysmobench, alhanahnah2025empirical, hasan2026automated} reflects the capability of LLMs in generating formal models and repairing them based on syntax errors and model checking results.
We adapted the Agent from SysMoBench~\cite{chengsysmobench} by removing the source code from the prompt, since the ground-truth implementation is not available in our problem setting.
Compared to the Basic Modeling Agent, PAT-Agent~\cite{zuo2025pat} includes an LLM-driven planning stage before building a formal model and employs Retrieval-Augmented Generation (RAG) with syntax documentation and in-context learning examples to improve modeling performance.
Event-B Agent~\cite{wang2026event} further introduces fine-grained verification and repair at the level of state transition events, and repairs both the model and the proofs based on proof obligation results.

\subsection{Metrics}

We adopt the metrics from SysMoBench~\cite{chengsysmobench} and Event-B Agent~\cite{wang2026event} to evaluate the quality of the generated formal models.
To evaluate syntactical performance, Syntax Correctness (Syntax) and Runtime Coverage (Runtime) metrics are adopted from SysMoBench, ensuring that the generated model is compilable and doesn't produce runtime errors during model checking.
The semantic performance of a formal model can be evaluated from various aspects.
Trace Conformance and Invariant Correctness from SysMoBench measure whether a formal system has the expected sequence of operations and whether each state is correct with respect to the invariants.
Event-B Agent proposes semantic metrics, including Proof Discharge Rate (PDR) and Requirement Fulfillment Rate (RF).
We adopt Trace Conformance (Trace), PDR, and RF as the semantic metrics, omitting Invariant Correctness.
This is because RF is based on proofs, which is stronger than Invariant Correctness through model checking.
In addition, because each requirement may involve multiple invariants, and the number of invariants generated depends on the LLM, RF ensures a fairer evaluation.

\subsection{RQ1. Overall Performance}

\input{tables/RQ1}

To answer RQ1, we compare \textit{ProGS} with existing agentic approaches for formal modeling.
Table~\ref{tab:rq1_overall} reports the overall performance of \textit{ProGS} and the baselines.
For each metric, the best result is shown in bold, and the second-best result is underlined.
\textit{ProGS} achieves either the best or second-best result across all metrics, demonstrating its effectiveness in constructing high-quality formal models with verifiable correctness.

The three baseline approaches are ordered by methodological complexity.
Basic Modeling Agent, which consists of one generation round followed by three repair iterations, most directly reflects the semantic capability of the underlying LLM.
It achieves a high Syntax Correctness score of 0.944, the highest Runtime Coverage score of 0.804, and the second-best Trace Conformance score of 0.629.
Together, these results indicate that even without a complex methodology, Basic Modeling Agent can generate compilable formal models that partially conform to the expected runtime behavior.
However, its Requirement Fulfillment Rate is only 0.778, leaving out 22.2\% of the invariants with at least one undischarged proof obligation.
This is relatively low given that automated provers and SMT solvers discharge 95.2\% of its proof obligations.
One possible explanation is that the generated models are semantically correct but contain specifications that existing tools struggle to verify automatically.
Another possibility is that the models contain semantic faults that are not exposed by model checking and therefore remain unrepaired by the Basic Modeling Agent.

Similarly, PAT-Agent uses model checking results to guide the repair of generated formal models.
Compared with Basic Modeling Agent, PAT-Agent introduces an additional planning stage and applies retrieval-augmented generation with syntax documentation and in-context learning examples.
As a result, it achieves the highest Syntax Correctness score of 0.981.
In addition, PAT-Agent derives formal verification properties from the requirements, including LTL properties and deadlock-freeness checks, and uses them to verify the generated formal model.
Its PDR and RF are comparable to those of Basic Modeling Agent.
However, PAT-Agent shows lower Runtime Coverage and Trace Conformance, mainly because it generates fixed constant values to initialize model checking and incorporates these values into the model itself.

Event-B Agent further introduces repairs guided by proof obligations, and allows both the formal model and the corresponding proofs to be revised.
It also supports model refinement in the pipeline to reduce modeling and verification effort at each abstraction level.
This approach substantially improves the Requirement Fulfillment rate to 0.941, suggesting that proof feedback can help LLMs identify and add missing model components.
However, modifying the model specification to discharge proof obligations may introduce semantic issues.
This is reflected in two ways.
First, Runtime Coverage and Trace Conformance decrease significantly, indicating that proof-oriented repairs can alter the model semantics and introduce unexpected deadlocks.
Second, the overall Proof Discharge Rate improves only marginally, partly because refinement increases the total number of proof obligations, including obligations for refinement correctness and the correctness of added lemmas or invariants.
The fact that some added components cannot be proven correct reflects that Event-B Agent may improve proof discharge at the cost of semantic performance.
This highlights the difficulty of repairing proofs while preserving the intended semantics of the original model.

In contrast to existing approaches, \textit{ProGS} achieves a more balanced performance across all evaluated aspects.
It maintains competitive Syntax Correctness, generates formal models with high proof-level verifiability in terms of PDR and RF, and preserves correct runtime semantics.
Overall, \textit{ProGS} consistently obtains the best or second-best result across all metrics, whereas each baseline exhibits substantial weaknesses in at least one dimension of proof-level or runtime performance.
In particular, \textit{ProGS} achieves the second-highest Requirement Fulfillment rate of 0.903 and the highest Proof Discharge Rate of 0.967.
Its PDR surpasses that of Event-B Agent, indicating that proof-guided repairs in \textit{ProGS} introduce fewer unverifiable components into the generated model.
In addition, the case splits encoded in model-based proof sketches reduce proof difficulty by decomposing complex proof obligations into simpler ones.
The RF of \textit{ProGS} is slightly lower than that of Event-B Agent because \textit{ProGS} keeps the environment model fixed after construction to support consistent structural checks.
This design prevents \textit{ProGS} from adding simple but useful auxiliary lemmas, such as $n \div 2 \in 0..n$, which leaves some proof obligations undischarged.
By contrast, Event-B Agent may silently add over-restrictive model components that discharge requirement-related proof obligations while leaving missing behavioral cases undetected, resulting in a higher RF but weaker semantic performance.

The runtime metrics further support this interpretation.
\textit{ProGS} achieves the highest Trace Conformance score of 0.697, outperforming PAT-Agent, Event-B Agent, and Basic Modeling Agent by 0.395, 0.210, and 0.068, respectively.
Compared with Basic Modeling Agent, whose semantic performance most directly reflects the capability of the underlying LLM, \textit{ProGS} achieves an even higher Trace Conformance score.
This suggests that the failure localization and structured repair guidance provided by \textit{ProGS} help the LLM improve the semantic correctness of the generated model.
The Runtime Coverage score of \textit{ProGS} is 0.777, slightly below the best result.
This is mainly due to limitations in the metric computation.
Although the structural checks in \textit{ProGS} enable the generated models to cover more scenarios than the baselines, Runtime Coverage is computed using only three traces for each system.
As a result, some state-transition events in the \textit{ProGS} models are not exercised during metric computation.
Moreover, some correctly modeled scenarios may be unreachable under the sampled traces.
For example, abnormal cases 4 and 6 in Figure~\ref{fig:motivation} implement error-handling behavior and are not generated by any baseline.


\subsection{RQ2. Efficiency and Modeling Efforts}

\input{tables/RQ2}
\input{tables/RQ2_modeling_effort}

To evaluate the efficiency of each method, we measure the average runtime in minutes (Avg. Time) and the average number of LLM requests (Avg. \#LLM Calls).
As shown in Table~\ref{tab:rq2_efficiency}, Basic Modeling Agent and PAT-Agent both complete formal model construction in less than five minutes on average and require fewer than four LLM requests.
This low cost reflects their limited repair process, where only a small number of faults can be detected and repaired within a few trials.
The low RF scores of both methods in Table~\ref{tab:rq1_overall} further indicate that model checking feedback alone is insufficient for constructing high-quality formal models that reliably satisfy the specified requirements.

In contrast, both Event-B Agent and \textit{ProGS} incorporate proof information into the feedback loop.
Because proof obligations are generated for each invariant-event pair, discharging all of them requires substantially more effort than model checking.
Table~\ref{tab:rq2_modeling_effort} reports the total number of proof obligations that cannot be automatically discharged by automated provers or SMT solvers, summed over all 27 examples in the dataset.
On average, each generated system has 4.89 and 4.44 undischarged POs for Event-B Agent and \textit{ProGS}, respectively, which is proportional to the amount of proof feedback provided to the LLM.
Event-B Agent repairs undischarged proof obligations individually, resulting in an average runtime of 25.98 minutes and 29.93 LLM requests in total.
After the repair rounds, 71 out of 132 undischarged POs remain unresolved, and Event-B Agent requires 7.148 LLM requests on average to discharge one PO.
These results illustrate the difficulty of repairing formal models using theorem-proving feedback.

\textit{ProGS} reduces modeling effort by using model-based proof sketches to structure proof feedback.
Failures can be localized to subtrees, which provide the LLM with more informative repair guidance than isolated proof obligations.
As a result, \textit{ProGS} requires substantially fewer LLM requests (2.366) to discharge each PO and resolves more proof obligations than Event-B Agent.
Its average runtime and number of LLM calls are also lower than those of Event-B Agent, despite the additional overhead of structural checks and the extra LLM requests used to identify dependencies among model components.

\subsection{RQ3. Ablation Study}

\input{tables/RQ3}

We evaluate the effectiveness of two key components of \textit{ProGS}, edge-level structural checks and subtree-based repair, through an ablation study.
Edge-level structural checks include deterministic normalization of generated model-based proof sketches and iterative repair based on the structural checks described in Section~\ref{sec:sketch}.
Subtree-based repair refers to localizing failures to subtrees and providing the LLM with aggregated subtree-level information to repair proof obligations associated with leaf nodes, as described in Section~\ref{sec:subtree_repair}.

We ablate these two features from the full version of \textit{ProGS} to construct three variants:
\begin{enumerate}
    \item \textbf{without (w.o.) Both Features}: The model-based proof sketch is generated by an LLM and then iteratively repaired using individual proof obligations as feedback, without structural checks or subtree-level repair.
    \item \textbf{without (w.o.) Subtree Repair}: The model-based proof sketch is cleaned and repaired using structural checks, but subsequent repair is guided only by individual proof obligations rather than subtree-level fault localization.
    \item \textbf{without (w.o.) Structural Checks}: Verification failures are localized to subtrees in the model-based proof sketch, and aggregated subtree-level information is provided to the LLM for iterative repair, but the sketch is not validated by structural checks.
\end{enumerate}
Table~\ref{tab:rq3} reports the performance of these ablated variants and the full version of \textit{ProGS}.
Overall, \textit{ProGS} achieves the best or second-best performance on all metrics except for Syntax Correctness.
Both features are demonstrated to be effective in improving the runtime or proving performance.

\begin{tcolorbox}[
    title=Finding 1,
    colback=gray!5,
    colframe=black,
    boxrule=0.5pt,
    arc=2pt
]
Edge-level structural checks improve the runtime performance.
\end{tcolorbox}

Edge-level structural checks help prevent LLM-based repairs from introducing unexpected deadlocks through over-restrictive model components.
They also ensure that child nodes remain semantically consistent with the conditions inherited from their parent nodes.

Comparing the \textit{w.o. Subtree Repair} baseline with the \textit{w.o. Both Features} baseline, we find that adding structural checks improves Runtime Coverage and Trace Conformance, although it slightly reduces proof-level performance.
A similar trend is observed when comparing the full version of \textit{ProGS} with the \textit{w.o. Structural Checks} baseline, except that the latter achieves a higher RF score.
These results indicate that edge-level structural checks are effective in preserving behavioral coverage and improving runtime semantics.
Although they may rule out over-restrictive shortcuts that make proof obligations easier to discharge, they help ensure that repaired sketches remain structurally valid and semantically consistent across parent-child relations.

\begin{tcolorbox}[
    title=Finding 2,
    colback=gray!5,
    colframe=black,
    boxrule=0.5pt,
    arc=2pt
]
Subtree-based repair improves both the proving and the runtime performance.
\end{tcolorbox}

Subtree-based repair localizes verification failures to regions of the model-based proof sketch that may share common root causes.
Since nodes within the same subtree share common guard conditions, failures to preserve the same invariant within that subtree may indicate a shared modeling fault.
Providing this aggregated subtree-level information to the LLM enables more coordinated and consistent repairs.

Adding subtree-based repair improves both proof-level and runtime performance.
This trend is observed by comparing the \textit{w.o. Both Features} baseline with the \textit{w.o. Structural Checks} baseline, and by comparing the \textit{w.o. Subtree Repair} baseline with the full version of \textit{ProGS}.
These results demonstrate the effectiveness of subtree-based repair in coordinating changes across related events and improving the overall quality of the generated formal models.

%% file: tables/RQ1.tex
\begin{table}[t]
\centering
\caption{Overall performance of \textit{ProGS} and the 3 baselines on 27 formal systems. The best result is shown in \textbf{bold}, and the second-best result is \underline{underlined}.}
\label{tab:rq1_overall}
\begin{tabular}{lccccc}
\toprule
\textbf{Method} & \textbf{Syntax} & \textbf{PDR} & \textbf{RF} & \textbf{Runtime} & \textbf{Trace} \\
\midrule
Basic Modeling Agent & \underline{0.944} & 0.952 & 0.778 & \textbf{0.804} & \underline{0.629} \\
PAT-Agent & \textbf{0.981} & 0.946 & 0.757 & 0.623 & 0.302 \\
Event-B Agent & 0.863 & \underline{0.956} & \textbf{0.941} & 0.617 & 0.487 \\
\textbf{ProGS} & \underline{0.944} & \textbf{0.967} & \underline{0.903} & \underline{0.777} & \textbf{0.697} \\
\bottomrule
\end{tabular}
\end{table}

%% file: tables/RQ2.tex
\begin{table}[t]
\centering
\caption{The average runtime and LLM calls used to construct a formal model.}
\label{tab:rq2_efficiency}
\begin{tabular}{lcc}
\toprule
\textbf{Method} & \textbf{Avg. Time (min)} & \textbf{Avg. \#LLM Calls} \\
\midrule
Basic Modeling Agent    & 3.73 & 2.37 \\
PAT-Agent     & 4.68 & 3.22 \\
Event-B Agent & 25.98 & 29.93 \\
\textbf{ProGS} & 18.43 & 21.59 \\
\bottomrule
\end{tabular}
\end{table}

%% file: tables/RQ2_modeling_effort.tex
\begin{table}[t]
\centering
\caption{Modeling effort during proof-guided repair. Calls/PO denotes the number of LLM calls required to discharge one previously undischarged proof obligation.}
\label{tab:rq2_modeling_effort}
\begin{tabular}{lccc}
\toprule
\textbf{Method} & \textbf{Undisch. Before} & \textbf{Undisch. After} & \textbf{Calls / PO} \\
\midrule
Event-B Agent & 132 & 71 & 7.148 \\
\textbf{ProGS} & 120 & 38 & 2.366 \\
\bottomrule
\end{tabular}
\end{table}

%% file: tables/RQ3.tex
\begin{table}[t]
\centering
\caption{RQ3 Ablation Studies.}
\label{tab:rq3}
\begin{tabular}{lccccc}
\toprule
\textbf{Method} & \textbf{Syntax} & \textbf{PDR} & \textbf{RF} & \textbf{Runtime} & \textbf{Trace} \\
\midrule
w.o. Both Features & 0.947 & 0.958 & 0.872 & 0.725 & 0.555 \\
w.o. Subtree Repair & \underline{0.948} & 0.953 & 0.860 & \underline{0.761} & \underline{0.608} \\
w.o. Structural Checks & \textbf{0.956} & \textbf{0.982} & \textbf{0.962} & 0.715 & 0.600 \\
\textbf{ProGS} & 0.944 & \underline{0.967} & \underline{0.903} & \textbf{0.777} & \textbf{0.697} \\
\bottomrule
\end{tabular}
\end{table}

%% file: 5.discussion.tex
\section{Discussion}

\subsection{Threats to Validity}

\subsubsection{Construct Validity}

A threat to construct validity is that proof discharge is used as a proxy for model verifiability. The correctness guarantees provided by \textit{ProGS} are bounded by the capabilities of the underlying verification tools.
Automated provers and SMT solvers may fail to discharge valid proof obligations that require domain-specific lemmas, suitable proof hints, or interactive reasoning. 
Consequently, an unproven obligation does not necessarily indicate an incorrect model.
Instead, it may reflect the incompleteness or limited automation of the prover. 
\textit{ProGS} should therefore be viewed as reducing, rather than eliminating, the need for expert intervention.

\subsubsection{Internal Threats}

A primary internal threat arises from the nondeterministic behavior of LLMs. 
Different runs may produce different model-based proof sketches and repair decisions, even when given the same input.
This randomness may affect both model quality and the number of synthesis or repair iterations required. 
Moreover, the measured performance may depend on prompt design, decoding parameters, and the selected LLM. 
We mitigate this threat by applying the same configurations and evaluation procedure to all methods. 
Nevertheless, repeated experiments with additional models and parameter settings would provide stronger evidence of result stability.



\subsubsection{External Threats}

Our benchmark provides partial environment information, such as the names and types of constants and variables.
While this constrains the space of formal models that LLMs may generate, it is necessary for computing Runtime Coverage and Trace Conformance using concrete constant assignments as inputs.
Because all evaluated methods receive the same requirement documents, the resulting restriction is applied uniformly and does not bias the comparison among methods.

%% file: 6.related_work.tex
\section{Related Work}

\subsection{Autoformalization}

Recent advances in LLMs have motivated growing interest in autoformalization within the formal methods community.
LLM-based approaches have been applied to tasks such as LLM-assisted theorem proving~\cite{jiang2023draft,zhao2024subgoal,wang2024lego,kasibatla2024cobblestone,zhou2025retrieval, zhou2025towards,yang2023leandojo}, specification generation~\cite{wen2024enchanting, endres2024can, wang2026tale,ma2025specgen}, and verifiable code generation~\cite{cai2025automated, mukherjee2025llm}, showing promising results.

Compared with these tasks, which target individual components of the formal-methods workflow, formal modeling involves a complete pipeline of model construction, verification, and validation, and iterative repair.
One line of work explores LLMs as interactive assistants for formal modeling.
For example, ModelWisdom~\cite{chen2026modelwisdom} integrates LLMs with TLA+ via a user interface to simplify the modeling process, while still leaving substantial modeling decisions to the user.
Another line of work empirically studies LLMs' ability to automatically construct formal models~\cite{hong2025effectiveness, alhanahnah2025empirical, mao2025llm}.
More recent agentic approaches further incorporate feedback from formal verification to guide model generation and repair~\cite{zuo2025pat,hasan2026automated,wang2026event,chengsysmobench,specula2025}.
Unlike these approaches, which generate and repair model components largely as a flat collection of artifacts, \textit{ProGS} introduces model-based proof sketches as an intermediate representation that structures the construction process, decomposes verification checks, and provides hierarchical guidance for repair.

\subsection{Theorem Proving with Proof Sketches}

Proof sketches and other intermediate representations have also been explored in LLM-assisted theorem proving.
These approaches typically decompose a target theorem into an informal proof draft or a sequence of subgoals before generating the corresponding formal proof~\cite{jiang2023draft, zhao2024subgoal, wang2024lego, kasibatla2024cobblestone, zhou2025retrieval, zhou2025towards}.
However, the sketches used in these works are primarily adopted as guidance for proof generation, rather than as intermediate model-based artifacts that are themselves verified and repaired.
As a result, they often provide a fixed decomposition of the proving task, with limited mechanisms for assessing or improving the sketch's quality during the procedure.
In contrast, our model-based proof sketches are designed for system-level formal model construction.
They encode not only proof subgoals, but also the state-transition events that discharge those subgoals, leading to a fundamentally different problem setting.

%% file: 7.conclusion.tex
\section{Conclusion}

Automating state-based formal modeling has long remained challenging because verification feedback is often reported at a granularity that does not directly reveal the underlying modeling faults.
We propose \textit{ProGS}, a new autoformalization method centered on model-based proof sketches, which serve as an intermediate representation for decomposing global verification into edge- and node-level checks, localizing faults precisely, and aggregating related failures through subtrees to guide repair.
Unlike existing approaches that treat generated formal models as flat collections of components, \textit{ProGS} structures the modeling process around hierarchical model-based proof sketches and produces high-quality formal models whose correctness can be more easily verified.
\textit{ProGS} offers a new paradigm for LLM-based formal modeling, where the model-based proof sketch creates opportunities to design new forms of structural and modular verification beyond those supported by flat formal models.
Future work can extend model-based proof sketches with additional edge- and subtree-level formal checks, and further investigate their relationship with formal model refinement and trace validation.

